\documentclass[pra,twocolumn,groupedaddress,showpacs,floatfix]{revtex4}

\usepackage{graphics}
\usepackage{graphicx}
\usepackage{bm}
\usepackage{amsmath}
\usepackage{amsfonts}
\usepackage{amssymb}
\usepackage{latexsym}
\usepackage{color}

\begin{document}

\title{A master equation for a two-sided optical cavity}
\author{Thomas M. Barlow, Robert Bennett, and Almut Beige}
\affiliation{The School of Physics and Astronomy, University of Leeds, Leeds LS2 9JT, United Kingdom}
\date{\today}

\begin{abstract}
Quantum optical systems, like trapped ions, are routinely described by master equations. The purpose of this paper is to introduce a master equation for two-sided optical cavities with spontaneous photon emission. To do so, we use the same notion of photons as in linear optics scattering theory and consider a continuum of traveling-wave cavity photon modes. Our model predicts the same stationary state photon emission rates for the different sides of a laser-driven optical cavity as classical theories. Moreover, it predicts the same time evolution of the total cavity photon number as the standard standing-wave description in experiments with resonant and near-resonant laser driving. The proposed resonator Hamiltonian can be used, for example, to analyse coherent cavity-fiber networks [Kyoseva {\em et al.}, New J. Phys. {\bf 14}, 023023 (2012)]. 
\end{abstract}
\pacs{42.50.Ex, 42.50.Pq}

\maketitle

\section{Introduction} \label{Intro}

It is a mathematical fact that any function on a finite interval can be written as a Fourier series. For example, any real-valued function $f(x)$ with $x \in (0,d)$ can be expanded in a series of exponentials,  
\begin{eqnarray} \label{FS}
f(x) &=& \sum_{m=-\infty}^\infty c_m \, \exp \left({\rm i} m \, {2 \pi x \over d} \right) \, ,
\end{eqnarray}
where the $c_m$ are complex coefficients with $c_m = c_{-m}^*$ \cite{Math}. This is usually taken as the starting point when quantising the electromagnetic field inside a perfect optical resonator or inside a dielectric slab or a so-called open cavity \cite{Loudonx,Knight,Abram,Knoell,Drummond,Glauber,Welsch,Khosravi,Barnett2,Dutra,Vivi,Scheel,Philbin}. Usually a finite quantisation volume is considered and the electromagnetic field observables are written as Fourier series of discrete sets of eigenfunctions. These eigenfunctions usually are the basic solutions of Maxwell's equations for the vector potential of the electromagnetic field in Coulomb gauge. The coefficients $c_m$ and $c_{-m}^*$ of these series are eventually replaced by photon annihilation and creation operators $c_m$ and $c_m^\dagger$, respectively. Subject to normalisation, the above-described canonical quantisation procedure yields a harmonic oscillator Hamiltonian of the form
\begin{eqnarray} \label{old}
H_{\rm cav} &=& \sum_{m=1}^\infty \hbar \omega_m \, c_m^\dagger c_m 
\end{eqnarray}
which sums over a discrete set of cavity frequencies $\omega_m$ (cf.~App.~\ref{single} for more details). Eq.~(\ref{old}) has been probed successfully experimentally with the help of single atoms passing through a resonator (cf.~eg.~Refs.~\cite{Walther,Haroche,experiments}). 

Nevertheless there is a problem. The standard Hamiltonian $H_{\rm cav}$ cannot be used to analyse other relatively straightforward experiments in a straightforward way. For example, suppose a monochromatic laser field of frequency $\omega_0$ drives a two-sided optical cavity from one side, thereby populating its normal modes. Moreover, suppose these modes are highly symmetric and couple equally well to the free radiation field on the left and on the right side of the resonator. Taking this point of view, one expects equal photon emission rates through both sides of the cavity. But this is not the case. Analysing a laser-driven optical resonator, a so-called Fabry-Perot cavity, with Maxwell's equations shows that resonant laser light is transmitted through the cavity with no reflected component (cf.~eg.~Ref.~\cite{Teich} or App.~\ref{Maxwell}). Off resonance, one part of the incoming laser beam is transmitted, while the other part is reflected. The corresponding transmission and reflection rates $T_{\rm cav}$ and $R_{\rm cav}$ are in general very different from each other.  

Of course, the above problem has been noticed before by other authors. Many different descriptions of the electromagnetic field between two mirrors exist in the literature. For example, taking a phenomenological approach, Collett and Gardiner \cite{Gardiner,Gardiner2} introduced the so-called input-output formalism. This formalism assumes a linear coupling between the photon modes outside and the photon modes inside the cavity and imposes boundary conditions for the electric field amplitudes on the mirrors. In this way, it becomes possible to model the coherent scattering of light through optical cavities in a way, which is consistent with Maxwell's equations (cf.~eg.~Refs.~\cite{Walls,Claire}). 

In addition to the input-output formalism, there are several modes-of-the-universe descriptions of optical cavities \cite{Scully,Ujihara,Geo-Benacloche,Dalton2,Dutra2}. These describe the electromagnetic field between two mirrors in terms of the normal modes of a much larger surrounding cavity, the universe. For example, Refs.~\cite{Dutra2,Dalton2} apply a macroscopic quantisation procedure to obtain a quasi-mode representation of the electromagnetic field. Quasi-modes are non-orthogonal photon modes. Hence tunneling between photon modes associated with the inside and the outside of the resonator can occur, thereby allowing for the leakage of photons through the cavity mirrors.

The purpose of this paper is to introduce an alternative model. In the following, we describe two-sided optical cavities with spontaneous photon emission by a quantum optical master equation. Master equations are routinely used to model laser-driven atomic systems, like trapped ions. As we shall see below, our approach is consistent with classical theories. Whether the input-output formalism, universe models, or quantum optical master equations describe optical cavities most accurately eventually has to be decided in the laboratory.

Before deriving our master equation, we notice that linear optics scattering theory and cavity quantum electrodynamics (QED) both employ different notions of photons. In cavity QED, photons are usually the energy quanta of the discrete modes of the electromagnetic field between two mirrors. In contrast to this, linear optics scattering theory only uses the term photon when referring to the energy quanta of free radiation fields. Resonator mirrors are usually seen as half-transparent mirrors which either transmit or reflect any incoming photon without changing its frequency. Since the mirrors affect their dynamics, the traveling-wave photons are in general different from the energy quanta of the electromagnetic field between two mirrors. 

In the following, we adopt the same notion of photons as in scattering theory. This means, we no longer use the mathematical argument sketched in Eq.~(\ref{FS}) to quantise the electromagnetic field between two mirrors. Instead we allow for a continuum of traveling-wave cavity photon modes. More concretely, we use the same Hilbert space when modelling the electromagnetic field inside an optical cavity and when modelling a free radiation field. In the following, $a_{\rm A} (\omega)$ denotes the annihilation operator of a photon with frequency $\omega$. The index ${\rm A} = {\rm L},{\rm R}$ helps to distinguish between left and right moving photons. For simplicity we restrict ourselves to only one polarisation degree of freedom. Photons in different $(\omega,{\rm A})$ modes are assumed to be in pairwise orthogonal states. Taking this approach makes it easy to guarantee that photons do not change their frequency when traveling through a resonator. Moreover, it allows us to assign different decay channels to photons traveling in different directions. It also enables us to assume that a laser which enters the cavity from the left excites only photons traveling right, as it should. A similar approach to optical cavities has recently been taken by Dilley {\em et al.}~\cite{Dilley}.  

The effect of the cavity mirrors is to convert photons traveling left into photons traveling right and vice versa until they eventually leak out of the resonator. This is in the following taken into account by postulating a cavity Hamiltonian $H_{\rm cav}$ with a coupling term that is known to be the generator of a unitary operation associated with the scattering of photons through beamsplitters and other linear optics elements \cite{Stenholm,Stenholm2,Leonhardt,Scheel}. Photons which are not in resonance with one of the cavity frequencies $\omega_m$ in Eq.~(\ref{old}) consequently experience level shifts. As pointed out by Glauber and Lewenstein \cite{Glauber}, photons and the energy quanta of an optical cavity seem to differ by some ``virtual" excitation. Only when the distance $d$ between the cavity mirrors tends to infinity, the coupling between photons travelling in different directions vanishes and the proposed cavity Hamiltonian simplifies to the usual free-space Hamiltonian.

The master equation which we derive in this paper contains two free parameters --- a coupling rate $J(\omega)$ and a spontaneous cavity decay rate $\kappa$. These can be chosen such that our model predicts the same stationary state light emission rates through the left and the right cavity mirror as Maxwell's equations. As we shall see below, both parameters depend on the photon round trip time. In addition, $J(\omega)$ depends on the amount of constructive and destructive interference within the cavity. The proposed master equation also predicts the same time-evolution of the total number of photons inside the cavity as the usual discrete-mode description for experiments with resonant and near-resonant laser driving. This means, the theory which we present here does not contradict already existing cavity QED experiments (cf.~eg.~Refs.~\cite{Walther,Haroche,experiments}). 

One advantage of the traveling-wave model which we propose here is that it makes it easy to analyse the spontaneous emission of photons through the different sides of an optical resonator or the scattering of photons through cascaded cavities \cite{Claudia,Peter}. It can also be used to describe the scattering of single photons through the fiber connections of coherent cavity networks. As long ago as 1997, Cirac {\em et al.}~\cite{Cirac} proposed a quantum internet by connecting distant optical cavities via very long optical fibers. In the mean time, much effort has been made to realise such schemes in the laboratory \cite{Kimble,Kuhn,Rempe}. Alternatively, cavities could be linked via fiber connections of intermediate length \cite{Pellizzari2,vanEnk,Zhou,Busch,Kyoseva}. For example, Kyoseva {\em et al.}~\cite{Kyoseva} proposed to create coherent cavity networks with very high or even complete connectivity by linking several cavities via linear optics elements and optical fibers, which are about 1m long. Using the approach which we propose here, it is relatively straightforward to analyse such networks analytically.

\begin{figure}[t]
\center
\includegraphics[width=8cm]{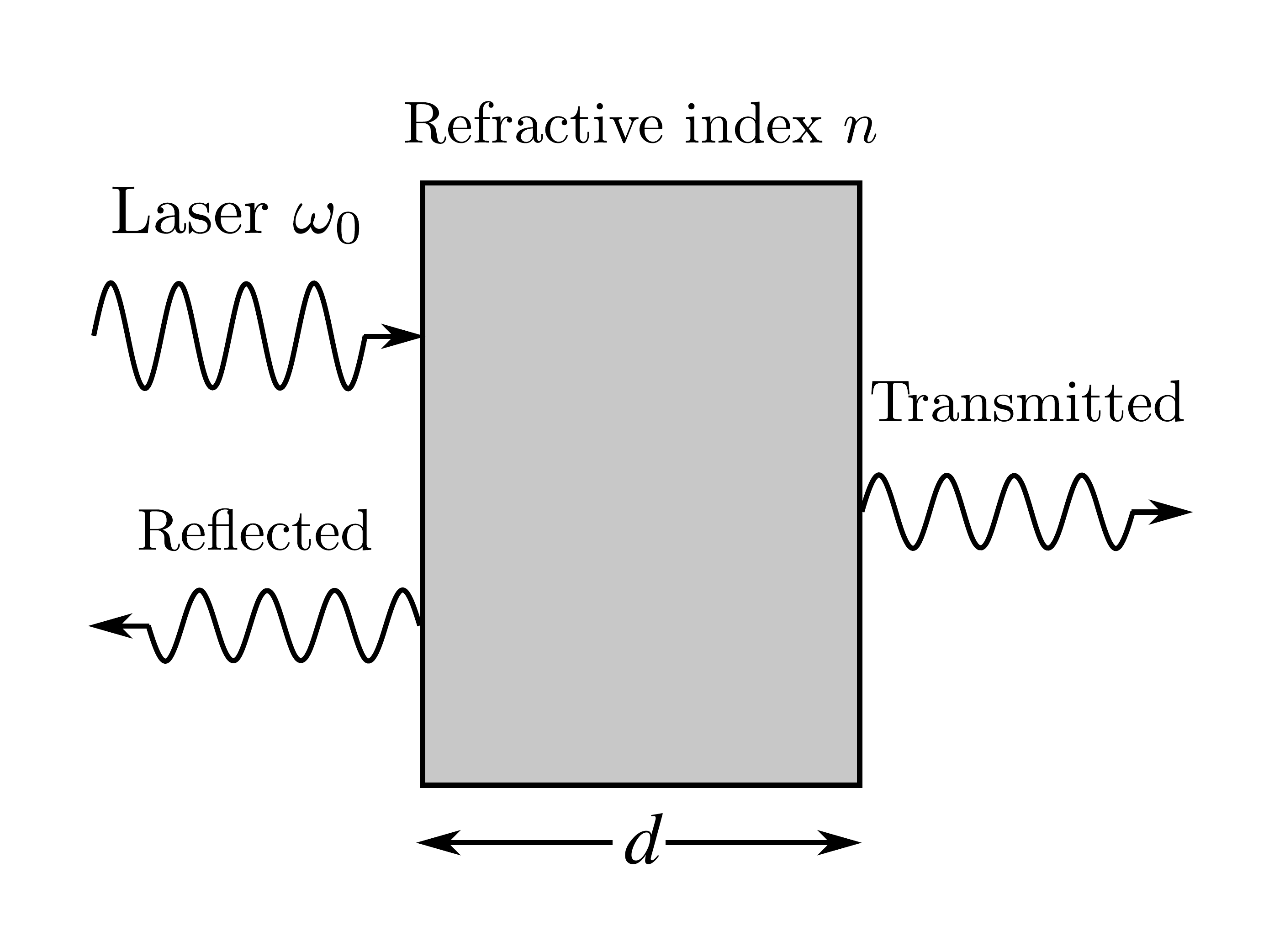} 
\caption{Schematic view of the experimental setup which we consider in this paper. It consists of a laser-driven resonator (a dielectric slab) of length $d$. Detectors monitor the stationary state photon emission rate through both cavity mirrors.} \label{fpcavity}
\end{figure}

There are five sections in this paper. Section \ref{two} postulates a traveling-wave Hamiltonian for two-sided optical cavities and introduces the corresponding master equation. Section \ref{sec3} uses this equation to calculate the stationary state photon scattering rates through the left hand and the right hand side of this experimental setup. Section \ref{ratesx} compares both rates with the stationary state scattering behaviour predicted by classical electrodynamics and obtains analytical expressions for the coupling constant $J(\omega)$ and the spontaneous cavity decay rate $\kappa$. Section \ref{ratesx} moreover shows that the standard description of optical cavities is consistent with our model for resonant and near resonant laser driving. In case of an infinitely long cavity, the proposed cavity Hamiltonian becomes the usually-assumed free field Hamiltonian. Finally, we summarise our findings in Section \ref{conclusions}. Apps.~\ref{single} and \ref{Maxwell} contain background material. 

\section{A traveling-wave cavity Hamiltonian} \label{two}

In this section, we introduce a traveling-wave description of the electromagnetic field inside an optical cavity. For simplicity we consider a so-called Fabry-P\'{e}rot or two-sided optical cavity (cf.~Fig.~\ref{fpcavity}) which consists of a dielectric slab of arbitrary length $d$ and has a refractive index $n >1$. An external monochromatic laser field with frequency $\omega_0$ drives the resonator from the left. The main reason for considering this relatively simple experimental setup is that its stationary state behaviour can be modelled easily with the help of Maxwell's equations (cf.~eg.~Ref.~\cite{Teich} and App.~\ref{Maxwell}), since absorption in the cavity mirrors remains negligible. Moreover, only a single polarisation, namely the polarisation of the applied laser field, needs to be taken into account. The generalisation of our results to arbitrary cavity designs is straightforward \cite{Tom2}. 

\subsection{Cavity photons}

In the following, we model the electromagnetic field inside a dielectric slab of a finite length $d$ using the same Hilbert space as when modelling an infinitely long slab. More concretely, we consider a continuum of photon modes with bosonic annihilation and creation operators $a_{\rm A}(\omega)$ and $a_{\rm A}^\dagger(\omega)$ with ${\rm A} = {\rm L}, {\rm R}$ and $\omega \in (0,\infty)$. In the following, we associate the corresponding photons with left and right-moving modes of frequency $\omega$. Photons in different modes are in general in pairwise orthogonal states. Annihilation and creation operators consequently obey the commutator relation
\begin{eqnarray} \label{comm}
[a_{\rm A} (\omega),a_{{\rm A}'}^\dagger (\omega')] &=& \delta_{{\rm A},{\rm A}'} \, \delta (\omega - \omega') 
\end{eqnarray}
with ${\rm A}, {\rm A}'= {\rm L}, {\rm R}$. The Hilbert space for the description of the dielectric slab in Fig.~\ref{fpcavity} contains all the states which are generated when applying the above photon creation operators to the vacuum state. However, different from an infinitely long dielectric slab, these photon modes exist only inside the cavity. 

Taking the same philosophy as linear optics scattering theory, the cavity mirrors become semi-transparent mirrors which transmit and reflect any incoming photon without changing its frequency. This frequency conservation suggests that a photon in the $a_{\rm R} (\omega)$-mode either remains in this mode or changes into the $a_{\rm L} (\omega)$-mode. This is in the following taken into account by writing the total Hamiltonian for the electromagnetic field inside the dielectric slab as  
\begin{eqnarray} \label{Hcav3}
H_{\rm cav} &=& H_{\rm field} + H_{\rm coup} \, ,
\end{eqnarray}
where $H_{\rm field}$ is the harmonic oscillator Hamiltonian  
\begin{eqnarray} \label{Hfield}
H_{\rm field} &=& \sum_{{\rm A} = {\rm L}, {\rm R}} \int_0^\infty {\rm d} \omega \, \hbar \omega \, a_{\rm A}^\dagger (\omega) a_{\rm A} (\omega)
\end{eqnarray}
which describes the free energy of the photons inside the resonator. Moreover, the coupling Hamiltonian 
\begin{eqnarray} \label{Hcav5}
H_{\rm coup} &=& {1 \over 2} \int_0^\infty {\rm d} \omega \, \hbar J (\omega) \, a_{\rm L}^\dagger (\omega) a_{\rm R} (\omega) + {\rm H.c.} 
\end{eqnarray}
describes the continuous conversion of photons traveling left into photons traveling right and vice versa with the (real) conversion rate $J(\omega)$.

The form of the above Hamiltonian might seem surprising, since it is usually assumed that a photon of frequency $\omega$ has the energy $\hbar \omega$. However, this applies only to the free field Hamiltonian $H_{\rm field}$ in Eq.~(\ref{Hfield}). When diagonalising $H_{\rm cav}$ in Eq.~(\ref{Hcav3}), we find that 
\begin{eqnarray} \label{Hcav500}
H_{\rm cav} &=& \int_0^\infty {\rm d} \omega \, \left( \hbar \omega + {1 \over 2} \hbar J (\omega) \right) \, a_+^\dagger (\omega) a_+ (\omega) \nonumber \\
&& + \left( \hbar \omega - {1 \over 2} \hbar J (\omega) \right) a_-^\dagger (\omega) a_- (\omega) \, , 
\end{eqnarray}
where the $a_\pm$, 
\begin{eqnarray} 
a_\pm &\equiv & {1 \over \sqrt{2}} \left( a_{\rm L} \pm a_{\rm R} \right) \, , 
\end{eqnarray}
denote standing-wave photon annihilation and creation operators. This means, the energy quanta of the electromagnetic field inside an optical cavity are its standing-wave photons. In the presence of the cavity mirrors, these standing wave photons can experience significant level shifts.

\subsection{Laser driving}

We now turn our attention to the experimental setup illustrated in Fig.~\ref{fpcavity}. In the presence of an external laser field, its Hamiltonian can be written as
\begin{eqnarray} \label{Hx}
H &=& H_{\rm cav} + H_{\rm laser}  
\end{eqnarray}
with the first term being the cavity Hamiltonian in Eq.~(\ref{Hcav3}) and with the second term taking the external laser driving into account. As in App.~\ref{single}, we treat the laser field classically. In addition, we assume that a laser which drives the cavity with frequency $\omega_0$ from the left only excites photons which are of the same frequency moving to the right. The interaction Hamiltonian for the coupling of laser light (from the left) into the cavity hence equals
\begin{eqnarray} \label{Hlaser2}
H_{\rm laser} &=& {1 \over 2} \hbar \Omega \, {\rm e}^{- {\rm i} \omega_0 t} \, a_{\rm R} ( \omega_0 ) + {\rm H. c.} 
\end{eqnarray}
in the Schr\"odinger picture. Notice that this Hamiltonian is the result of the presence of a laser field inside the cavity. The laser Rabi frequency $\Omega$ in Eq.~(\ref{Hlaser2}) is therefore a direct measure for laser amplitude inside (but not outside) the resonator \footnote{This is different from the input-output formalism, where the laser Rabi frequency $\Omega$ is a measure for the laser amplitude outside the cavity.}.

As long as only a single laser field with frequency $\omega_0$ is applied, only photons in the $a_{\rm L}(\omega_0)$ and in the $a_{\rm R}(\omega_0)$ mode become populated eventually. All other photon modes can be ignored. Ignoring in addition the frequency dependence of constants and operators, when it is obvious, and introducing the interaction picture with respect to 
\begin{eqnarray}
H_0 &=&  \sum_{{\rm A} = {\rm L}, {\rm R}} \hbar \omega_0 \, a_{\rm A}^\dagger a_{\rm A} \, , 
\end{eqnarray}
the Hamiltonian $H$ in Eq.~(\ref{Hx}) simplifies to the interaction Hamiltonian 
\begin{eqnarray} \label{H100}
H_{\rm I} &=& {1 \over 2} \hbar \Omega \left( a_{\rm R} + a_{\rm R}^\dagger \right)  
+  {1 \over 2} \hbar J \left( a_{\rm L}^\dagger a_{\rm R} + {\rm H.c.} \right) \, . ~~~~
\end{eqnarray}
We now have a time-independent Hamiltonian to describe a laser-driven two-sided optical cavity. 

\subsection{Cavity leakage} \label{leakagesec}

In order to take the possible leakage of photons through the resonator mirrors into account, we add a system-bath interaction term to the above Hamiltonian and then trace out the bath-degrees of freedom on a coarse grained time scale $\Delta t$ \cite{Knight}. Since we distinguish between left and right moving photons, it is now straightforward to assign different decay channels to photons traveling in different directions. Cavity photons in the $a_{\rm R}$-mode leave the cavity through the right mirror. Analogously, photons in the $a_{\rm L}$-mode only leak out through the left mirror. In the following, we denote the corresponding spontaneous decay rate by $\kappa$. This decay rate is the same for left and right moving photons due to the symmetry of the experimental setup in Fig.~\ref{fpcavity}. 

If we describe the system in Fig.~\ref{fpcavity} by a density matrix $\rho_{\rm I}$, then the corresponding left and right photon emission rates $I_{\rm A}$ are given by 
\begin{eqnarray} \label{IX}
I_{\rm A} &=& \kappa \, {\rm Tr} \left( a_{\rm A}^\dagger a_{\rm A} \rho_{\rm I} \right) 
\end{eqnarray}
with ${\rm A} ={\rm R}, {\rm L}$. In other words, the photon emission probability density is the mean number of photons in the $a_{\rm A}$-mode multiplied with $\kappa$. The quantum optical master equation of Lindblad form which reflects this emission behaviour is given by
\begin{eqnarray} \label{master2}
\dot \rho_{\rm I} &=& - {{\rm i} \over \hbar} \left[ H_{\rm I} , \rho_{\rm I} \right] + \sum_{\rm A=L,R} {1 \over 2} \kappa \, \big( 2 a_{\rm A} \rho_{\rm I} a_{\rm A}^\dagger \nonumber \\
&& - a_{\rm A}^\dagger a_{\rm A} \rho_{\rm I} - \rho_{\rm I} a_{\rm A}^\dagger a_{\rm A}  \big) \, .  ~~
\end{eqnarray}
In the following, we use this equation to analyse the dynamics of the laser-driven optical cavity.

\section{The time evolution of photon number expectation values} \label{sec3}

In this section, we calculate the stationary state photon emission rates $I_{\rm L}^{\rm ss}$ and $I_{\rm R}^{\rm ss}$ through the left and the right cavity mirror, respectively. The sum of these is the total photon emission rate
\begin{eqnarray} \label{sumx2}
I_{\rm Tot}^{\rm ss} &\equiv & I^{\rm ss}_{\rm L} + I^{\rm ss}_{\rm R} \, .
\end{eqnarray}
To calculate these rates we use rate equations, ie.~linear differential equation which describe the time evolution of expectation values. 

\subsection{Time evolution of expectation values}

To obtain the relevant rate equations, we notice that the above master equation can be used to show that the expectation value $\langle A_{\rm I} \rangle$ of an observable $A_{\rm I}$ in the interaction picture evolve according to the differential equation
\begin{eqnarray} \label{dotA02}
\langle \dot A_{\rm I} \rangle &=& -{{\rm i} \over \hbar} \, \left\langle\left[A_{\rm I},H_{\rm I} \right]\right\rangle + \sum_{\rm A=L,R} {1 \over 2} \kappa \, \langle 2 a_{\rm A}^\dagger A_{\rm I} a_{\rm A} \nonumber \\
&& - A_{\rm I} a_{\rm A}^\dagger a_{\rm A} - a_{\rm A}^\dagger a_{\rm A} A_{\rm I} \rangle \, . ~~~
\end{eqnarray}
To find a closed set of rate equations, including equations for the time evolution of the mean photon number in the $a_{\rm L}$ and in the $a_{\rm R}$ mode, respectively, we need to consider the expectation values 
\begin{eqnarray}
&& \hspace*{-0.5cm} n_{\rm L} \equiv \langle a_{\rm L} ^\dagger a_{\rm L} \rangle \, , ~~ 
n_{\rm R} \equiv \langle a_{\rm R} ^\dagger a_{\rm R} \rangle \, ,  \nonumber \\
&& \hspace*{-0.5cm} k_1 \equiv \langle a_{\rm L} + a_{\rm L}^\dagger \rangle \, , ~~ 
k_2 \equiv  {\rm i} \langle a_{\rm R} - a_{\rm R}^\dagger \rangle \, , \nonumber \\
&& \hspace*{-0.5cm} k_3 \equiv {\rm i} \langle a_{\rm L} a_{\rm R}^\dagger - a_{\rm L}^\dagger a_{\rm R} \rangle \, . ~~
\end{eqnarray}
These five variables evolve according to the linear differential equations
\begin{eqnarray} \label{rateeq}
\dot n_{\rm L} &=& {1 \over 2} J k_3 - \kappa n_{\rm L} \, , ~~ \nonumber \\
\dot n_{\rm R} &=& {1 \over 2} \Omega k_2  - {1 \over 2} J k_3 - \kappa n_{\rm R} \, , ~~ \nonumber \\
\dot k_1 &=& - {1 \over 2} J k_2 - {1 \over 2} \kappa k_1 \, , ~~ \nonumber \\
\dot k_2 &=& \Omega + {1 \over 2} J k_1 - {1 \over 2} \kappa k_2 \, , ~~ \nonumber \\
\dot k_3 &=& - \Omega k_1 - J (n_{\rm L} - n_{\rm R}) - \kappa k_3 ~~~
\end{eqnarray}
which form a closed set. 
 
\subsection{Photon scattering rates}
  
Using Eq.~(\ref{IX}), one can now show that the photon emission rate $I_{\rm A}$ with ${\rm A} = {\rm L}, {\rm R}$ is simply given by 
\begin{eqnarray} \label{17}
I_{\rm A} &=& \kappa n_{\rm A} \, .
\end{eqnarray}
Proceeding as in App.~\ref{single} and setting all time derivatives equal to zero, we obtain the stationary state photon numbers
\begin{eqnarray} \label{numbers}
n_{\rm L}^{\rm ss} = {\Omega^2 J^2 \over \left( J^2 + \kappa^2 \right)^2} \, , ~~
n_{\rm R}^{\rm ss} = {\Omega^2 \kappa^2 \over \left( J^2 + \kappa^2 \right)^2} \, .
\end{eqnarray}
Substituting these into Eq.~(\ref{17}) yields different stationary state photon emission rates for the different sides of a laser-driven resonator,
\begin{eqnarray} \label{numbersx}
I_{\rm L}^{\rm ss}  = {\Omega^2 J^2 \kappa \over \left( J^2 + \kappa^2 \right)^2} \, , ~~
I_{\rm R}^{\rm ss} = {\Omega^2 \kappa^3 \over \left( J^2 + \kappa^2 \right)^2} \, .
\end{eqnarray}
The total stationary state photon emission rate $I_{\rm Tot}^{\rm ss}$ in Eq.~(\ref{sumx2}) hence equals
\begin{eqnarray} \label{21}
I_{\rm Tot}^{\rm ss} &=& {\Omega^2 \kappa \over J^2 + \kappa^2} \, .
\end{eqnarray} 
One can easily check that $I_{\rm Tot} = \kappa n_{\rm Tot}$ with $n_{\rm Tot} \equiv n_{\rm L} + n_{\rm R}$. This means the total emission rate depends only on the total cavity photon number, as it should. 

\subsection{Time evolution without laser driving} \label{nodriving2}

Before we compare the above photon emission rates with the predictions of classical electrodynamics, we consider the case when there is no external laser driving. When $\Omega = 0$, then one can show that the time derivative of the total number of cavity photons $n_{\rm Tot}$ equals
\begin{eqnarray} \label{rateeq2}
\dot n_{\rm Tot} &=& - \kappa n_{\rm Tot} 
\end{eqnarray}
without any approximations.  

\section{Consistency of quantum and classical models} \label{ratesx}

In the following, we compare the above predictions of a quantum-optical master equation with the predictions of classical scattering theory in App.~\ref{Maxwell} to find out, how the spontaneous cavity decay rate $\kappa$ and the photon coupling rate $J$ depend on the frequency $\omega$ of the respective photon modes, the length of the dielectric slab $d$, and its refractive index $n$. As we shall see below, $\kappa$ and $J$ are both a function of the photon round trip time. In addition, the coupling rate $J$ depends on the amount of constructive and destructive interference within the cavity. This section also discusses the consistency of the derived master equation for a two-sided optical cavity with alternative quantum optics models. It is shown that for near resonant laser driving, our model predicts exactly the same dynamics for the total number of photons inside the cavity as the standard single-mode description and is therefore in good agreement with existing quantum optics experiments \cite{Walther,Haroche,experiments}.

\subsection{Consistency with classical electrodynamics} \label{conditions}

Below we list several conditions which guarantee the consistency between our traveling-wave master equation and the predictions of classical electrodynamics (cf.~App.~\ref{Maxwell}): 
\begin{enumerate}
\item In the case of no laser driving, both models should predict the same relative flux of energy out of the cavity. Using the same notation as in Sections \ref{nodriving} and \ref{nodriving2}, this condition applies when
\begin{eqnarray} \label{sumx}
{\dot I (t) \over I(t)} &=& {\dot n(t) \over n(t)} \, .
\end{eqnarray}
\item In the case of laser driving, the stationary state photon emission rates $I^{\rm ss}_{\rm L}$ and $I^{\rm ss}_{\rm R}$ should have the same dependence on $\omega_0$, $d$, and $n$ as the classical cavity reflection and transmission rates $R_{\rm cav}(\omega_0)$ and $T_{\rm cav}(\omega_0)$. More concretely, we want that
\begin{align} \label{cond1}
\frac{I_L^{\rm ss}}{I_{\rm Tot}^{\rm ss}} = R_{\rm cav}(\omega_0) \, , ~~
\frac{I_R^{\rm ss}}{I_{\rm Tot}^{\rm ss}} &= T_{\rm cav}(\omega_0) \, .
\end{align}
The ratios on the right hand sides of these equations should not depend on the laser Rabi frequency $\Omega$, since there is no $\Omega$ in the classical model.
\end{enumerate}
In the following, we use the above conditions, to determine the two constants $\kappa$ and $J$ which we introduced in Section \ref{two}. 

For example, substituting Eqs.~(\ref{I(t)}) and (\ref{rateeq2}) into Eq.~(\ref{sumx}), we find that the energy flux equality condition applies when  
\begin{eqnarray} \label{kappa}
\kappa &=& - {2 c \over n d} \, \ln r \, . 
\end{eqnarray}
In this equation, $r$ is the Fresnel coefficient in Eq.~(\ref{B4}) for the reflection of photons from the dielectric back into the dielectric. The logarithm of $r$ guarantees that $\kappa = 0$ for $r=1$. This means, for perfectly reflecting mirrors, light stays forever inside the cavity. When $r \to 0$, then $\kappa \to \infty$ and there is effectively no cavity. 

To obtain an explicit expression for the coupling rate $J$, we now have a closer look at condition 2. Combining Eqs.~(\ref{numbersx})--(\ref{21}), one can easily show that
\begin{align} \label{cond1zzz}
\frac{I_L^{\rm ss}}{I_{\rm Tot}^{\rm ss}} = {J^2 \over J^2 + \kappa^2} \, , ~~
\frac{I_R^{\rm ss}}{I_{\rm Tot}^{\rm ss}} = {\kappa^2 \over J^2 + \kappa^2} \, .
\end{align}
Comparing these two equations with Eq.~(\ref{FP}), and using the above result for $\kappa$, we find that
\begin{align} \label{JJJ}
J (\omega_0) &= {4 c \over n d} \cdot {r \ln r \over 1 - r^2} \, \sin \left( \omega_0 {n d \over c} \right)
\end{align}
up to an overall phase factor. The coupling rate $J$ contains an interference term, which implies that photons of certain frequencies are more likely to be reflected by the cavity mirrors than others. For example, for resonant laser light, ie.~for a laser with an $\omega_0$ which is equal to one of the frequencies $\omega_m$ in Eq.~(\ref{omegam}), the photon coupling rate $J(\omega_0)$ becomes zero. This means, our model correctly predicts that resonant light does not get reflected within the cavity. 

Finally, let us consider the special case of highly-reflecting cavity mirrors. In this case, the Fresnel coefficient $r$ is very close to one. Hence 
\begin{eqnarray}
- 2 \ln r &=& 1 - r^2
\end{eqnarray}
to a very good approximation and Eqs.~(\ref{kappa}) and (\ref{JJJ}) simplify to 
\begin{eqnarray} \label{kappax}
\kappa &=& {c \over n d} \, (1-r^2) \, , \nonumber \\
J (\omega_0) &=& - {2 r c \over n d} \, \sin \left( \omega_0 {n d \over c} \right) \, .
\end{eqnarray}
The spontaneous decay rate $\kappa$ of a two sided optical cavity and the photon coupling rate $J$ depend only on the relative resonator length $d$, its refractive index $n$, and the frequency $\omega_0$ of the incoming light.

\subsection{Consistency with the standard single-mode description for near-resonant laser driving} \label{conquan}

The previous subsection shows that the constants $J$ and $\kappa$ of our traveling-wave master equation for a two-sided optical cavity can be adjusted such that it is consistent with the predictions of classical electrodynamics. However, there is already a well-established standing-wave model for optical cavities with external laser driving (cf.~App.~\ref{single} for more details). The purpose of this subsection is to show that our model is moreover consistent with the predictions of this model, at least for resonant and for near-resonant laser driving. This means, our traveling-wave cavity Hamiltonian does not contradict already existing quantum optics experiments which probe the field inside an optical cavity with the help of atomic systems (cf.~eg.~Ref.~\cite{Walther,Haroche,experiments}).  

\subsubsection{Resonant cavities}

When the laser is on resonance, ie.~when $\omega_0$ equals one of the frequencies $\omega_m$ in Eq.~(\ref{omegam}), then $J$ in Eq.~(\ref{JJJ}) becomes zero,
\begin{eqnarray}
J (\omega_0) &=& 0 \, .
\end{eqnarray}
This means, there is effectively no coupling between left and right travelling photons due to interference effects. For example, $n_{\rm L}$ remains zero, when the laser field populates only right moving photon modes. Using Eq.~(\ref{rateeq}), one can indeed show that  
\begin{eqnarray} \label{rateeq3}
\dot n_{\rm L} &=& - \kappa n_{\rm L} 
\end{eqnarray}
in this case. Under these conditions, there is a relatively simple closed set of rate equations which describe the time evolution of $n_{\rm R}$. Eq.~(\ref{rateeq}) shows that
\begin{eqnarray} \label{rateeq2xxx}
\dot n_{\rm R} &=& {1 \over 2} \Omega k_2 - \kappa n_{\rm R} \, , ~~ \nonumber \\
\dot k_2 &=& \Omega - {1 \over 2} \kappa k_2 
\end{eqnarray}
without any approximations. Consequently, the stationary state photon emission rates $I^{\rm ss}_{\rm L}$, $I^{\rm ss}_{\rm R}$, and $I^{\rm ss}_{\rm Tot}$ 
are given by
\begin{eqnarray}
I^{\rm ss}_{\rm L} = 0 ~~ {\rm and} ~~
I^{\rm ss}_{\rm R} = I_{\rm Tot}^{\rm ss} = {\Omega^2 \over \kappa} \, .  
\end{eqnarray}
This means, the total stationary state photon emission rate $I_{\rm Tot}^{\rm ss}$ is exactly the same as the one we obtain when using the quantum optical standard standing-wave description in App.~\ref{single}. We only need to identify the single-mode photon number $n$ with $n_{\rm R}$ and set the detuning $\Delta$ in Eq.~(\ref{Iss}) equal to zero. 

\subsubsection{Near-resonant cavities}

As we shall see below, the standard single-mode description of optical cavities also holds to a very good approximation for near-resonant laser driving, if we are only interested in the time evolution of the total cavity photon number $n_{\rm Tot}$. To do so, we notice that the photon coupling rate $J$ in Eq.~(\ref{kappax}) for near-resonant laser driving is to a very good approximation given by
\begin{eqnarray} \label{JDelta}
J &=& - 2 \Delta \, ,
\end{eqnarray}
as long as the cavity mirrors are highly-reflecting and the Fresnel coefficient $r$ is close to unity. Here $\Delta$ equals the detuning $\Delta_m$ in Eq.~(\ref{Deltam}) of the applied laser field from the nearest cavity resonance $\omega_m$. 

Taking this and Eq.~(\ref{rateeq}) into account, we moreover notice that a closed set of rate equations for the time evolution of $n_{\rm Tot}$ is given by
\begin{eqnarray} \label{rateeq3zzz}
\dot n_{\rm Tot} &=& {1 \over 2} \Omega k_2 - \kappa n_{\rm Tot} \, , ~~ \nonumber \\
\dot k_1 &=& - \Delta k_2 - {1 \over 2} \kappa k_1 \, , ~~ \nonumber \\
\dot k_2 &=& \Omega + \Delta k_1 - {1 \over 2} \kappa k_2 \, .
\end{eqnarray}
These equations are exactly the same as the rate equations in Eq.~(\ref{rates}), if we replace the single-mode photon number $n$ by the total photon number $n_{\rm Tot}$ of the model which we propose in this paper. In other words, the single mode description in App.~\ref{single} correctly predicts the total photon emission rate $I_{\rm Tot}^{\rm ss}$ of a laser-driven optical cavity. In agreement with Eq.~(\ref{Iss}), it equals
\begin{eqnarray} \label{lorentz}
I_{\rm Tot}^{\rm ss} &=& {\Omega^2 \kappa \over 4 \Delta^2 + \kappa^2} 
\end{eqnarray}
which is a Lorentzian function of $\Delta$. However, the standard standing wave description of optical cavities cannot predict the stationary state photon emissions rate through the different sides of two-sided cavities. In contrast to this, our standing-wave description of optical optical cavities (cf.~Eq.~(\ref{numbersx})) predicts that
\begin{eqnarray} \label{numbersz}
I_{\rm L}^{\rm ss} = {4 \Omega^2 \Delta^2 \kappa \over \left( 4 \Delta^2 + \kappa^2 \right)^2} \, , ~~
I_{\rm R}^{\rm ss} = {\Omega^2 \kappa^3 \over \left( 4 \Delta^2 + \kappa^2 \right)^2} 
\end{eqnarray}
for near-resonant laser driving. 

\subsubsection{The free radiation field} \label{freefield}

Finally, let us have a closer look at the case where the distance $d$ of the cavity mirrors tends to infinity. From Eqs.~(\ref{kappa}) and (\ref{JJJ}) we immediately see that 
\begin{eqnarray}
\kappa = J (\omega_0) = 0 
\end{eqnarray}
in this case. This is exactly as one would expect. If the resonator is infinitely long, then its photons remain inside forever and never change their direction. One can easily check that $J \equiv 0$ reduces the cavity Hamiltonian $H_{\rm cav}$ in Eq.~(\ref{Hcav3}) to the free field Hamiltonian $H_{\rm field}$ in Eq.~(\ref{Hfield}), by construction.

\section{Conclusions} \label{conclusions}

There is a close analogy between excited atomic systems and excited optical cavities. In both cases, a detector placed some distance away from the source registers spontaneously emitted photons. Like atoms, optical cavities have a spontaneous decay rate, which is usually denoted by $\kappa$. Atomic systems with spontaneous photon emission are routinely described by quantum optical master equations. The main result of this paper is the justification of such a master equation for a laser-driven two-sided optical cavity, which allows us to distinguish between photons leaking out through the left and through the right side of the resonator. To obtain such a master equation, we postulate the cavity Hamiltonian $H_{\rm cav}$ in Eq.~(\ref{Hcav3}). It allows us to assign different decay channels to photons travelling in different directions and guarantees that photons do not change their frequency when traveling through a cavity. 

The cavity Hamiltonian $H_{\rm cav}$ needs to be postulated such that its predictions are consistent with those of classical physics, whenever both theories apply. To justify its validity, we therefore apply it to a situation which can also be analysed by taking a fully classical approach. We assume that a two-sided optical cavity is driven by a monochromatic laser field with frequency $\omega_0$ (cf.~Fig.~\ref{fpcavity}). We then calculate the intensity of the transmitted and of the reflected light using either classical electrodynamics (cf.~App.~\ref{Maxwell}) or a quantum optical master equation which derives from Eq.~(\ref{Hcav3}). Both models are shown to yield the same stationary state reflection and transmission rates, if we choose the cavity decay rate $\kappa$ and the photon coupling rate $J$ as suggested in Eqs.~(\ref{kappa}) and (\ref{JJJ}).  

The cavity Hamiltonian $H_{\rm cav}$ in Eq.~(\ref{Hcav3}) acts on a distinct, large Hilbert space with a continuum of photon frequencies $\omega$, which is usually only considered when modelling free radiation fields. As in free space, we distinguish left and right moving modes. In this way, we find that it becomes possible to assume that a laser field which enters the setup from the left excites only photons traveling right, as it should. The cavity decay rate $\kappa$ for the leakage of photons through either side of the cavity depends, as one would expect, on the refractive index $n$ and the length $d$ of the dielectric slab (cf.~Eq.~(\ref{kappa})). The effect of the cavity mirrors is to change the direction of photons inside the resonator. They convert left into right moving photons and vice versa. The corresponding photon coupling rate $J$ in Eq.~(\ref{JJJ}) depends, like $\kappa$, on $n$ and $d$ but also on the laser frequency $\omega_0$, thereby accounting for the amount of constructive and destructive interference within the resonator.

As predicted by Maxwell's equations, there is no conversion of photons when the cavity is resonantly driven by an applied laser field. In this case, $J$ in Eq.~(\ref{JJJ}) becomes zero. For near resonant laser driving, $J$ becomes identical to $- 2 \Delta$ with $\Delta $ being the respective laser detuning. In this case one can show that the total cavity photon number $n_{\rm Tot}$ evolves in the same way as the photon number $n$ in the usually assumed single-mode standing-wave description of optical cavities (cf.~App.~\ref{single}). This means, the cavity theory which we propose here does not contradict current cavity QED experiments which probe the electromagnetic field inside an optical resonator with the help of atomic systems (cf.~eg.~Refs.~\cite{Walther,Haroche,experiments}). But now that a new cavity Hamiltonian is established, it can be used to describe physical scenarios which are beyond the scope of classical electrodynamics. For example, the proposed master equation can be used to describe cascaded cavities \cite{Claudia,Peter} and the scattering of single photons through the fiber connections of coherent cavity networks with
complete connectivity \cite{Kyoseva}. 

Our approach might be criticised for being phenomenological instead of deriving its equations via a rigorous field quantisation method, like macroscopic QED. The same criticism has previously been applied to the input-output formalism. A wealth of work has been done to reconcile various cavity QED theories (cf.~eg.~Refs.~\cite{Dutra,compare,Philbin}). However, macroscopic QED still contains several ad-hoc assumptions. It is not as rigorous as it might appear, since quantum physics does not tell us, which Hilbert space to choose, how to define photons in a gauge-independent way, and how to implement boundary conditions. For example, in this paper, we implement boundary conditions by choosing constants such that the stationary state of the laser-driven two-sided cavity is consistent with Maxwell's equations. But we do not restrict the Hilbert space in which photons live. More experiments are needed to decide which theory describes optical cavities most accurately. \\[0.5cm]
\noindent {\em Acknowledgements.} T.~B.~acknowledges financial support from a White Rose Studentship Network on Optimising Quantum Processes and Quantum Devices for future Digital Economy. R.~B.~thanks the UK Engineering and Physical Sciences Research Council for an EPSRC Doctoral Prize. Moreover, we thank Joshua Ezekiel Sambo, Anthony Hayes, Axel Kuhn, and Stefan Weigert for useful and stimulating discussions. 

\appendix
\section{Predictions of the standard standing-wave cavity Hamiltonian} \label{single}

In this appendix, we review the standard standing-wave description of the electromagnetic field between two mirrors and examine some of its predictions. As we shall see below, this model is only well suited for the description of the time evolution of the total number of photons inside an optical cavity with resonant or near-resonant laser driving. 

\subsection{The cavity-laser Hamiltonian}

In the standard model, the Hamiltonian of the experimental setup in Fig.~\ref{fpcavity} is of the general form
\begin{eqnarray} \label{H}
H &=& H_{\rm cav} + H_{\rm laser} \, .
\end{eqnarray}
The first term describes the free energy of the electromagnetic field inside the resonator. The second term takes the external laser driving into account. When quantising the electromagnetic field in the way of most textbooks, one derives at the assumption that the field only contains standing-wave photon modes of frequency $\omega_m$ with 
\begin{eqnarray} \label{omegam}
\omega_m &=& m \pi {c \over nd} \, ,
\end{eqnarray}
where $m$ is a positive integer, $c$ is the speed of light, $n$ is the refractive index of the medium inside the cavity, and $d$ is the distance of the resonator mirrors. If $c_m$ is the corresponding photon annihilation operator, $H_{\rm cav}$ simply equals the Hamiltonian in Eq.~(\ref{old}). The laser field is usually treated as a classical field. Denoting its Rabi frequencies by $\Omega_n$ and by frequency $\omega_0$, its Hamiltonian equals 
\begin{eqnarray} \label{Hlaser}
H_{\rm laser} &=& \sum_{m=1}^\infty {1 \over 2} \hbar \Omega_m \, {\rm e}^{- {\rm i} \omega_0 t} \, c_m + {\rm H. c.} 
\end{eqnarray}
This Hamiltonian arises from a spatial overlap of the classical driving field and the quantised field in the vicinity of the cavity mirrors.

When changing into the interaction picture with respect to the free Hamiltonian $H_0 = \sum_{m=1}^\infty \hbar \omega_0 \, c_m^\dagger c_m$ and after applying the usual rotating-wave approximation, we obtain the time-independent interaction Hamiltonian
\begin{eqnarray}
H_{\rm I} &=& \sum_{m=1}^\infty {1 \over 2} \hbar \Omega_m \left( c_m + c_m^\dagger \right) + \hbar \Delta_m \, c_m^\dagger c_m  
\end{eqnarray}
with the cavity-laser detuning $\Delta_m$ defined such that
\begin{eqnarray} \label{Deltam}
\Delta_m &\equiv & \omega_m - \omega_0 \, . 
\end{eqnarray}
For simplicity, we assume in the following that the frequency $\omega_0$ is relatively close to only one of cavity resonance frequencies $\omega_m$. Then only one of the cavity modes has to be taken into account and $H_{\rm I}$ simplifies to 
\begin{eqnarray}
H_{\rm I} &=& \hbar \Omega \left( c + c^\dagger \right) + \hbar \Delta \, c^\dagger c \, ,  
\end{eqnarray}
after neglecting the respective index $m$ for operators and constants. This Hamiltonian is often used in the literature when describing a laser-driven optical cavity. However, notice that this model does not distinguish whether the laser drives the cavity from the left or from the right. Here the laser only excites a single standing-wave photon mode.

\subsection{The corresponding master equation}

The spontaneous leakage of photons through the cavity mirrors is in the following taken into account via the usual quantum-optical master equation 
\begin{eqnarray} \label{master}
\dot \rho_{\rm I} &=& - {{\rm i} \over \hbar} \left[ H_{\rm I} , \rho_{\rm I} \right] + {1 \over 2} \kappa \left( \, 2 c \rho_{\rm I} c^\dagger - c^\dagger c \rho_{\rm I} - \rho_{\rm I} c^\dagger c \, \right) ,  ~~~~
\end{eqnarray}
where $\kappa $ is the cavity decay rate and $\rho_{\rm I}$ denotes the density matrix of the quantised cavity field. This equation can be derived by coupling the $c$-mode to a continuum of free radiation field modes outside the cavity, letting the system evolve over a short time $\Delta t$, and tracing out the free radiation field to mimic the effects of a photon-absorbing environment. Using the above standing-wave description, it is not possible to assign different decay channels to photons traveling in different directions.  
  
\subsection{Time evolution of expectation values}

The most straightforward way of calculating the intensity of the emitted light is to adopt a rate equation approach. Taking into account that the expectation value of any observable $A_{\rm I}$ in the interaction picture equals $\langle A_{\rm I} \rangle = {\rm Tr} (A_{\rm I} \rho_{\rm I})$, we find that 
\begin{eqnarray} \label{dotA0}
\langle \dot A_{\rm I} \rangle &=& - {{\rm i} \over \hbar} \, \left\langle\left[A_{\rm I},H_{\rm I} \right]\right\rangle - {1 \over 2} \kappa \, \langle A_{\rm I} c^\dagger c + c^\dagger c A_{\rm I} - 2 c^\dagger A_{\rm I} c \rangle \, . \nonumber \\
\end{eqnarray}
Here we are especially interested in the time evolution of the mean photon number $n$, 
\begin{eqnarray}
n &\equiv & \langle c^\dagger c \rangle \, . 
\end{eqnarray}
In order to obtain a closed set of rate equations, including one for $n$, we also need to consider the expectation values
\begin{eqnarray}
k_1 \equiv \langle c + c^\dagger \rangle \, , ~~ 
k_2 \equiv  {\rm i} \langle c - c^\dagger \rangle \, .
\end{eqnarray}
Using Eq.~(\ref{dotA0}), one can then show that $n$, $k_1$, and $k_2$ evolve according to the linear differential equations
\begin{eqnarray} \label{rates}
\dot n &=& {1 \over 2} \Omega k_2 - \kappa n \, , ~~ \nonumber \\
\dot k_1 &=& - \Delta k_2 - {1 \over 2} \kappa k_1 \, , ~~ \nonumber \\
\dot k_2 &=& \Omega + \Delta k_1 - {1 \over 2} \kappa k_2 \, .
\end{eqnarray}

\subsection{Stationary state photon emission rate}

To obtain expressions for the stationary state of the laser-driven resonator, we simply set the time derivatives in Eq.~(\ref{rates}) equal to zero. Doing so, we find for example that the stationary state cavity photon number $n^{\rm ss}$ equals
\begin{eqnarray} \label{nsingle}
n^{\rm ss} &=& {\Omega^2 \over 4 \Delta^2 + \kappa^2} \, .
\end{eqnarray}
The corresponding stationary state photon emission rates equals $I^{\rm ss} = \kappa n^{\rm ss} $ which implies
\begin{eqnarray} \label{Iss}
I^{\rm ss} &=& {\Omega^2 \kappa \over 4 \Delta^2 + \kappa^2} \, .
\end{eqnarray}
As we shall see in Section \ref{conquan}, this emission rate describes the leakage of photons through the left {\em and} the right cavity mirror.

\section{Predictions of classical scattering theory} \label{Maxwell}

Consider the experimental setup in Fig.~\ref{fpcavity} of a dielectric slab of width $d$ and refractive index $n$. If we assume normal incidence and consider the idealized case, where we can ignore diffraction, we can treat the system as one-dimensional. In the following, we denote the permittivity of the dielectric slab by $\varepsilon(x)$ such that 
\begin{eqnarray}
\varepsilon(x) &=& \left\{  \begin{array}{ll} n^2 ~~ & {\rm for} ~ x \in (0,L) \, , \\ 1 & {\rm elsewhere} \, , \end{array} \right.
\end{eqnarray}
if the mirror surfaces are placed at $x=0$ and $x=d$. Moreover, $\mu(x)$ is the permeability of the dielectrics. For simplicity we assume in the following $\mu(x) \equiv 1$.

\subsection{Continuous laser driving}

Suppose monochromatic laser light with frequency $\omega_0$ and wave vector $k_0 = \omega_0/c$ enters the Fabry-P\'{e}rot cavity in Fig.~\ref{fpcavity}. One can then use the standard Fresnel coefficients for radiation incident from a vacuum region upon a dielectric of refractive index $n$,
\begin{eqnarray}
r' = {1-n \over 1+n} ~~ {\rm and} ~~ t' = {2 \over 1+n} \, ,
\end{eqnarray}
and those for the same physical situation but with the incident wave from the other direction,
\begin{eqnarray} \label{B4}
r = {n-1 \over n+1} ~~ {\rm and} ~~ t = {2n \over n+1} \, ,
\end{eqnarray}
to write the relative amplitude of the electric field which leaves the cavity after having travelled $m$ times across as
\begin{eqnarray}
E_{\rm T} (x,m)  &=& t' \, r^{m-1} \, {\rm e}^{{\rm i} m k_0 nd} \, t \, .
\end{eqnarray} 
Here $x=0$, when $m$ is even and $x=d$, when $m$ is odd, since light that is ultimately reflected back into the direction of the incoming laser beam has even $m$ and light that is transmitted has odd $m$. The above equation takes into account that the electric field amplitude accumulates a phase factor ${\rm e}^{{\rm i} n k_0 d}$ every time it propagates the length $d$ of the cavity. 

\begin{figure}[t]
\center
\includegraphics[width=\columnwidth]{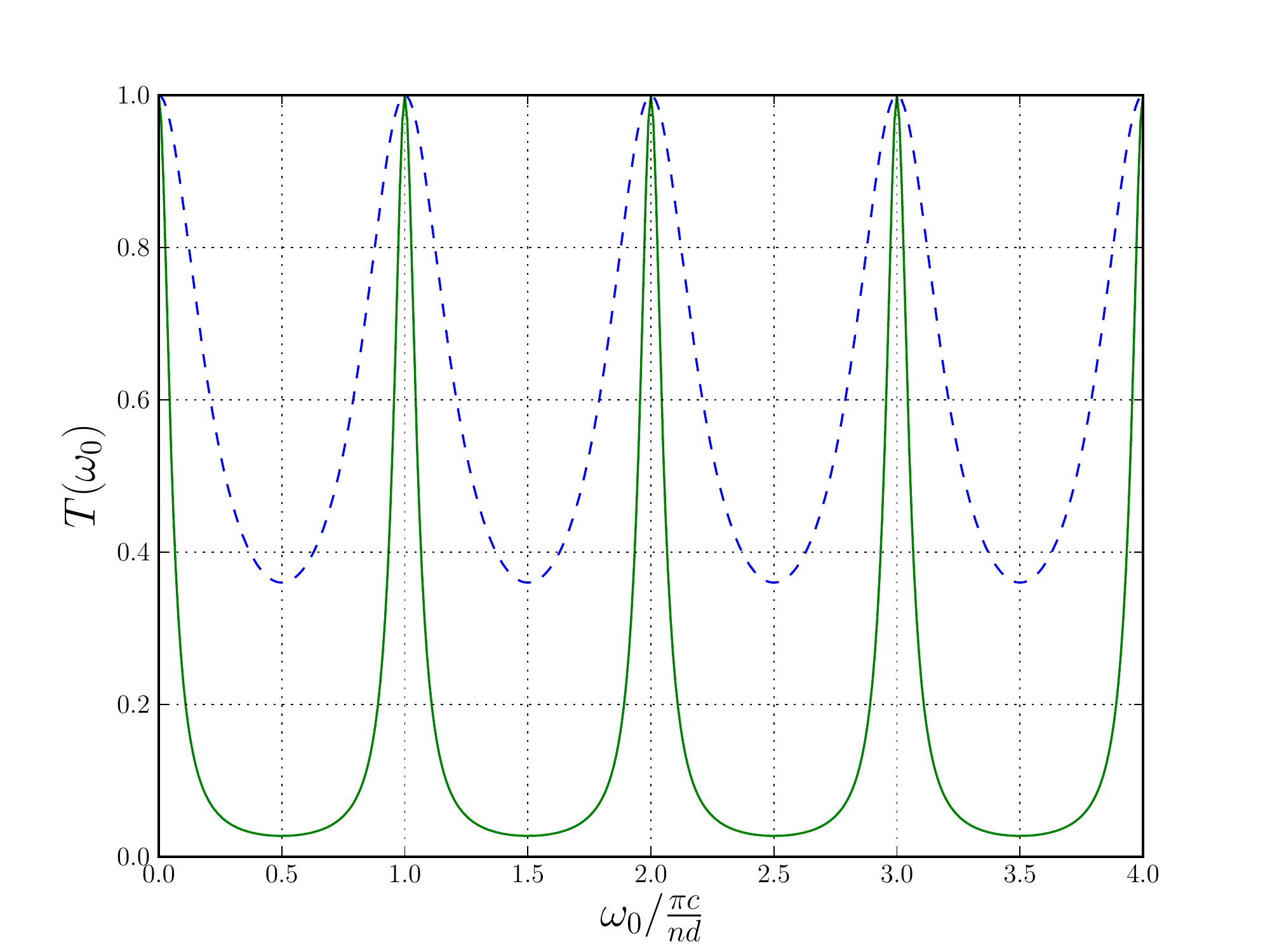}
\caption{Transmission rate $T_{\rm cav} (\omega_0)$ in Eq.~(\ref{FP}) of a Fabry-P\'{e}rot cavity which is driven by monochromatic light of frequency $\omega_0$ for the refractive index $n = 3$ (dashed line) and $n=20$ (solid line).} \label{transmission}
\end{figure}

The electric field of the reflected light also has a contribution of $r'$ from the component of the light that does not enter the cavity. The total reflection and transmission coefficients of the Fabry-P\'{e}rot cavity for normal incidence are therefore given by
\begin{align}
r_{\rm cav}(\omega_0) &= r' + \sum\limits_{m \rm{~even}} E_{\rm T} (0,m) \, , \nonumber \\
t_{\rm cav}(\omega_0) &= \sum\limits_{m \rm{~odd}} E_{\rm T} (d,m) 
\end{align}
which implies
\begin{align}
r_{\rm cav}(\omega_0) &= r' + t' \sum\limits_{m \rm{~even}} r^{m-1} \, {\rm e}^{{\rm i} m k_0 nd} \, t \, , \nonumber \\
t_{\rm cav}(\omega_0) &= t' \sum\limits_{m \rm{~odd}} r^{m-1} \, {\rm e}^{{\rm i}m k_0 nd} \, t \, .
\end{align}
When calculating these geometric series, we obtain
\begin{align}
r_{\rm cav}(\omega_0) &= r \, \frac{{\rm e}^{2{\rm i} k_0 nd}-1}{1-r^2 \, {\rm e}^{2{\rm i} k_0 nd}} \, , \nonumber \\
t_{\rm cav}(\omega_0) &= \frac{1-r^2}{1-r^2 \, {\rm e}^{2{\rm i} k_0 nd}} \, {\rm e}^{{\rm i} k_0 nd} \, .
\end{align}
The overall cavity reflection and transmission rates $R_{\rm cav}(\omega_0)$ and $T_{\rm cav}(\omega_0)$ are given by the modulus squared of the corresponding relative amplitudes. Hence $R_{\rm cav}(\omega_0) = |r_{\rm cav}|^2$ and $T_{\rm cav}(\omega_0) = |t_{\rm cav}|^2$ which implies
\begin{align} \label{FP}
R_{\rm cav}(\omega_0) &= \frac{F \sin^2 (k_0 nd)}{1 + F\sin^2 (k_0 nd)} \, , \nonumber \\
T_{\rm cav}(\omega_0) &= \frac{1}{1 + F \sin^2 (k_0 nd)} 
\end{align}
with $r$ as in Eq.~(\ref{B4}). The factor 
\begin{eqnarray}
F &=& {4r^2 \over (1-r^2)^2} 
\end{eqnarray}
is known as the finesse of the cavity.

Fig.~\ref{transmission} illustrates the dependence of the relative amplitude of the transmitted light on its frequency $\omega_0$ and on the refractive index $n$. As usual, we see that laser light with a frequency equal to one of the cavity resonance frequencies $\omega_m$ in Eq.~(\ref{omegam}) does not get reflected by the cavity. This means, resonant light travels through the resonator, as if it were not there. In general, we find that the larger the refractive index $n$, the more the light is affected by the dielectric. For relatively large $n$, there is almost complete reflection for some frequencies $\omega_0$. For $n$ close to 1, traveling through the dielectric is almost like traveling through the vacuum. In this paper, we seek a quantum master equation approach to optical cavities that reproduces these amplitudes.

\subsection{Time evolution without laser driving} \label{nodriving}

Suppose no external laser field is applied and a single wave packet bounces back and forth inside the two-sided cavity which is shown in Fig.~\ref{fpcavity}. This wave packet is a superposition of plane waves. Again, we assume that all waves in the packet experience the same refractive index $n$, so that all parts of the wave packet travel with the same speed. After $m$ bounces, the intensity of the wave at a fixed frequency $\omega_0$ equals
\begin{eqnarray}
I( t_m ) &=& r^{2 (m-1)} \, I(0) \, ,
\end{eqnarray}
where $I (0)$ is the initial intensity of the wave and 
\begin{eqnarray}
t_m &\equiv & m {n d \over c} 
\end{eqnarray}
is the time it takes a wave packet to bounce $m$ times through a medium of length $d$ and with refractive index $n$. To simplify a later comparison with the predictions of a quantum model, we notice that the intensity 
\begin{eqnarray} \label{I(t)}
I( t ) &=& r^{2 c t / n d} \, I(0)
\end{eqnarray}
assumes exactly the same value as $I(t_m)$ for $t=t_m$.

\end{document}